# On the Theory of Vibronic Superradiance

## A. P. Saiko

*Institute of Solid State Physics and Semiconductors, Belarussian Academy of Sciences, Minsk, 220072 Belarus*
*e-mail: saiko@ifttp.bas-net.by*

**Abstract**—The Dicke superradiance on vibronic transitions of impurity crystals is considered. It is shown that parameters of the superradiance (duration and intensity of the superradiance pulse and delay times) on each vibronic transition depend on the strength of coupling of electronic states with the intramolecular impurity vibration (responsible for the vibronic structure of the optical spectrum in the form of vibrational replicas of the pure electronic line) and on the crystal temperature through the Debye–Waller factor of the lattice vibrations. Theoretical estimates of the ratios of the time delays, as well as of the superradiance pulse intensities for different vibronic transitions well agree with the results of experimental observations of two-color superradiance in the polar dielectric $KCl:O_2^-$. In addition, the theory describes qualitatively correctly the critical temperature dependence of the superradiance effect.

## INTRODUCTION

A system of two-level particles, transferred, initially, into an excited state, due to interaction with vacuum oscillations of the radiation field, drops to the ground state with emission of an electromagnetic pulse. If the dephasing is inefficient, the dipole moments of individual particles, in the process of spontaneous emission, become correlated, i.e., a macroscopic polarization arises in the system, which is proportional to the number of inverted particles $N$. As a result, the intensity of the emitted pulse appears to be quadratic in $N$, with its duration being inversely proportional to $N$. This kind of decay of the inverted system is referred to as spontaneous coherent emission or merely superradiance. This phenomenon was first predicted by Dicke [1] (for the history of the problem and references see [2–5]). Superradiance differs from steady-state laser emission and from the amplified spontaneous emission proceeding within the time intervals comparable with the dephasing time in that the emission intensity, in the two latter processes, is proportional to the density of the inverted particles.

Experimentally, the superradiance was mostly observed in gases [6], though in a less pure form it was also detected in impurity crystals, namely, in polar dielectrics [7, 8] and aromatic compounds [3, 9–14]. These experiments, as well as the implementation of laser action in the region of the phonon wing of the optical spectra of crystals (see, e.g., special issues on solid-state lasers [15, 16]), indicate the key role of electron–phonon coupling in the formation of the coherent states of the field and matter. It is important that emission of an ensemble of optical impurities in crystal is usually regarded as a pure electronic transition between the excited and ground states. In this approach, the effect of the lattice or intramolecular vibrations on the impurity particle is weak and is mainly reduced to a shift and broadening of the spectral line of the optical transition. However, for sufficiently strong interaction of the optical centers with the lattice and intramolecular vibrations, the elementary events of absorption and emission of the light quanta are accompanied by creation and annihilation of vibrational quanta, which gives rise to formation of the electronic–vibrational (vibronic) structure of the optical spectra and, thus, complicates the processes of formation of both the laser emission and superradiance. Therefore, the phenomena under consideration should acquire a number of features not inherent in the usual situation when the effect of the environment on the electronic transition is weak. In particular, the emission of the superradiance pulses and laser emission may become possible on vibronic, rather than only on zero-phonon, transitions (on the phonon sideband), with the characteristic parameters of the phenomena being dependent on the strength of the electron–phonon coupling and on the crystal temperature. The problem of taking into account the strong electron–phonon coupling for laser action in a solid-state gain medium was first discussed by McCumber in [17]. Within the framework of a phenomenological approach based on the principle of detailed balance (which expresses in a generalized form the relation between the Einstein $A$ and $B$ coefficients) McCumber has calculated the gain factor for the vibronic laser as a function of the light frequency, populations of electronic levels, and crystal lattice temperature. In these calculations, the microscopic nature and form of the electron–phonon coupling were not taken into account. Only a few papers are available where the role of the strong electron–phonon coupling, upon formation of laser emission and superradiance is analyzed from the microscopic point of view (see, e.g., [3, 9–14, 18–29]).

The development of the microscopic theory of superradiance and laser emission in a resonance crystal medium with appropriate allowance for the strength and form of electron–phonon coupling, as well as specific features of the vibrational spectrum and temperature of the crystal lattice, is important not only for understanding of the physical processes and correct interpretation of experimental data, but also for the elaboration of principles of optimal selection of the materials in which the above phenomena may be implemented.

The superradiance in impurity crystals was analyzed from the microscopic point of view in [3, 9–14] for the case of weak coupling between emitters and phonons and in [18–21] for the case of strong coupling. Specific features of the superradiance within the framework of Thompson's model [22], which describes the processes of Brillouin scattering in resonance media with emission or absorption of a single phonon, were considered in [24, 25]. A phenomenological approach with the use of three- and four-level energy diagrams of optical impurities in crystals was developed in [26, 27].

In what follows, we propose a consistent microscopic theory of the vibronic superradiance oriented to a real experimental situation of observation of this phenomenon on vibronic transitions of the $O_2^-$ molecules in the polar dielectrics KCl:$O_2^-$ [7, 8].

## FORMULATION OF THE MODEL

In the KCl:$O_2^-$ system, the electronic transition is strongly coupled to one of the high-frequency intramolecular vibrations and much more weakly to other lattice vibrations. The frequency of the intramolecular mode exceeds several times the limiting frequencies of the lattice modes. According to the general theory [30–32], the luminescence (or optical absorption) spectrum should consist, in this case, of a series of narrow well-resolved lines representing replicas of the pure electronic line related to intramolecular vibrations. Each line of the series should be accompanied by the phonon wing, arising due to excitation of crystal vibrations together with the intramolecular ones.

For the density matrix $\rho$ of the impurity system interacting with the electromagnetic field, we may write the master equation

$$\frac{\partial \rho(t)}{\partial t} = -i(L_f + L_a + L_{af} + L_l + L_{al} + L_L \quad (1)$$
$$+ L_{aL} + i\Lambda_f + i\Lambda_a + i\Lambda_l)\rho(t),$$

where

$$L_m X \equiv [H_m, X], \quad (2)$$

$H_m$ and $L_m$ are the Hamiltonians and the corresponding Liouvillians ($m = a, f, l$, etc.); $H_f = \omega a^+ a$ and $H_a = \varepsilon \sum_j R_j^z$ are the Hamiltonians of the single-mode electromagnetic field (the emission is collinear to the axis of the rod-like sample) and optical two-level centers; $H_{af} = g\sum_j (aR_j^+ + \text{H.c.})$ is the operator of interaction between the optical centers and radiation field; $H_l = \nu b^+ b$ is the Hamiltonian of the intramolecular vibration; $H_{al} = \sum_j \xi \nu R_j^z (b + b^+)$ is the electron–vibrational interaction responsible for the vibronic structure of the optical spectrum; $H_L = \sum_k \omega_k c_k^+ c_k$ is the lattice Hamiltonian; $H_{aL} = \sum_k \lambda_k R_j^z (c_k + c_k^+)$ is the operator of interaction of the impurity with the lattice modes; $a^+$, $b^+$, $c_k^+$ and $\omega$, $\nu$, $\omega_k$ are the operators of creation of the light field quanta, intramolecular vibrations, $k$th lattice mode, and the relevant frequencies; $g$, $\xi$, and $\lambda_k$ are the coupling coefficients; $R_j^{\pm, z}$ are the energy spin operators [33, 34], which describe the $j$th two-level optical center ($j = 1, 2, \ldots, N$) and are identical to the Pauli spin matrices; and the Planck constant is everywhere assumed to be unity. The Liouvillians $\Lambda_f$ and $\Lambda_a$ take into account incoherent interactions that lead to energy dissipation of the light field and excited optical impurities, respectively:

$$\Lambda_f X = \text{k}([aX, a^+] + \text{э.c.}) + 2\kappa\bar{N}[a, [X, a^+]], \quad (3)$$

$$\Lambda_a X = \frac{1}{2}\sum_j \{\gamma_{12}([R_j^- X, R_j^+] + \text{H.c.}) \quad (4)$$
$$+ \gamma_{21}([R_j^+ X, R_j^-] + \text{H.c.})\},$$

where k is the decay constant of the light field due to the irreversible escape of photons beyond the bounds of the prolate sample with the length $l$ (we may assume that k = $c/l$, where $c$ is the speed of light); $\bar{N} = [\exp(\omega/k_B T) - 1]^{-1}$, $T$ is the sample temperature, and $\gamma_{12}(\gamma_{21})$ is the rate of transition from the ground (excited) electronic state of the impurity to the excited (ground) state. The Liouvillian $\Lambda_l$ describes the decay of the intramolecular vibration:

$$\Lambda_l X = \alpha([bX, b^+] + \text{H.c.}) + 2\alpha\bar{n}[b, [X, b^+]], \quad (5)$$

where $\alpha$ is the decay constant and $n = [\exp(\nu/k_B T) - 1]^{-1}$.

Information on the form and derivation of Liouvillians (3)–(5) that allow for dissipative processes may be found, e.g., in [35–37].

## ELIMINATION OF THE FAST VARIABLES

In the problem of superradiance, the polarization ($\sim R_j^\pm$) and inversion ($\sim R_j^z$) of optical impurities are slow variables, whereas the variables of the radiation field, as well as those of the local vibrations and lattice modes, are fast variables. Therefore, the last-mentioned may be adiabatically separated using general methods of nonequilibrium statistical mechanics [37] and, thus, the master equation for the density matrix of the ensemble of impurities may be obtained. The sought equation of the second order in the interaction $H_{af}$ (or $L_{af}$) may be represented in the form of a differential (rather than integro-differential, as is commonly done) equation, by using the method developed in [38]:

$$\frac{\partial \sigma(t)}{\partial t} = -\int_0^t dt' \langle \tilde{L}_{af}(t)\tilde{L}_{af}(t')\rangle_{f,l,L}\sigma(t) + \Lambda_a\sigma(t), \quad (6)$$

where

$$\tilde{L}_{af}(t) = V(t)L_{af}V^+(t), \quad (7)$$

$$V(t) = \exp[i(L_a + L_f + L_l + L_L + L_{al} + L_{aL} + i\Lambda_f + i\Lambda_l)t]. \quad (8)$$

Note that the density matrix of the impurity system, with allowance for the transformation (8), is defined as

$$\sigma(t) = \langle V(t)\rho(t)\rangle_{f,l,L}, \quad (9)$$

where $\langle \ldots \rangle_{f,l,L} = \mathrm{Sp}_f\mathrm{Sp}_l\mathrm{Sp}_L\{\ldots \rho_f\rho_l\rho_L\}$, $\rho_{l,L} = \exp(-H_{l,L}/k_BT)/\mathrm{Sp}_{l,L}[\exp(-H_{l,L}/k_BT)]$ are the density matrices of the local and lattice modes, $\rho_f = |0\rangle\langle 0|$ is the density matrix of the radiation field (for the frequencies in the optical range the electromagnetic field may be assumed to be in the vacuum state $|0\rangle$ at zero temperature). We also take into account that $\langle \tilde{L}_{af}(t)\rangle_{f,l,L} = 0$.

After substituting proper expressions into Eq. (6) and after successive "disentanglement" of all the commutation relations, with allowance for definitions (2)–(5), and after evaluating the traces, we have

$$\frac{\partial \sigma(t)}{\partial t} = -g^2 \mathrm{Sp}_{l,L}\left\{\int_0^{t'} dt'[e^{-i\omega(t-t')}e^{-k|t-t'|}(\tilde{R}^+(t)\tilde{R}^-(t')\right.$$
$$\left. \times \sigma(t)\rho_l\rho_L - \tilde{R}^-(t')\sigma(t)\rho_l\rho_L\tilde{R}^+(t)) + \mathrm{H.c.}]\right\} + \Lambda_a\sigma(t), \quad (10)$$

where the collective variables $R^{\pm,z} = \sum_j R_j^{\pm,z}$ are introduced and, according to (7), $\tilde{R}^\pm = VR^\pm V^+$. By differentiating with respect to time, one can easily make sure that the operator $\tilde{R}^-(t)$ obeys the equation

$$\frac{\partial \tilde{R}^-(t)}{\partial t} = -i\varepsilon\tilde{R}^-(t) - iF(t)\tilde{R}^-(t), \quad (11)$$

where

$$F(t) = V(t)\left[\xi\nu(b+b^+) + \sum_k \lambda_k(c_k+c_k^+)\right]V^+(t). \quad (12)$$

A similar equation for $\tilde{R}^+$ may be obtained directly from Eq. (11) using Hermitian conjugation. As a result, we have

$$\tilde{R}^-(t) = e^{-i\varepsilon t}\overset{(-)}{T}\left(\exp\left(-i\int_0^t F(t')dt'\right)\right)R^- \equiv e^{-i\varepsilon t}U(t)R^-, \quad (13)$$

$$\tilde{R}^+(t) = e^{i\varepsilon t}R^+\overset{(+)}{T}\left(\exp\left(i\int_0^t F(t')dt'\right)\right) \equiv e^{i\varepsilon t}R^+U(t), \quad (14)$$

where $\overset{(-)}{T}$ and $\overset{(+)}{T}$ are the operators of chronological and antichronological ordering.

By substituting Eqs. (13) and (14) into Eq. (10), we obtain

$$\frac{\partial \sigma(t)}{\partial t} = -g^2\int_0^t dt'[e^{-i(\omega-\varepsilon)(t-t')}e^{-k|t-t'|}\langle U^+(t)U(t')\rangle_{l,L} \quad (15)$$
$$\times (R^+R^-\sigma(t) - R^-\sigma(t)R^+) + \mathrm{H.c.}] + \Lambda_a\sigma(t).$$

In the calculation of the correlation function

$$\langle U^+(t)U(t')\rangle_{l,L}$$
$$= \left\langle \overset{(+)}{T}\left(\exp\left(i\int_0^t dt_1 F(t_1)\right)\right)\overset{(-)}{T}\left(\exp\left(-i\int_0^{t'} dt_2 F(t_2)\right)\right)\right\rangle_{l,L} \quad (16)$$

we retain in the operator $F$ only the first-order terms in the coupling coefficients $\xi$ and $\lambda_k$, thus eliminating the dependence of this operator on the energy spin variable $R^z$. Now, we may represent correlation function (16) in the form of a cumulant expansion and calculate it in the same way as was done in a similar case in [35, 36]. Due to the linearity of the interaction operators $H_{al}$ and $H_{aL}$ with respect to bosonic variables, the expansion terminates at the second cumulant and we have

$$\langle U^+(t)U(t')\rangle_{l,L} = \exp\left[-\int_0^t dt_1\int_0^{t_1} dt_2 \langle F(t_2)F(t_1)\rangle_{l,L}\right.$$
$$-\int_0^{t'} dt_1\int_0^{t_1} dt_2 \langle F(t_1)F(t_2)\rangle_{l,L} \quad (17)$$
$$\left.+\int_0^t dt_1\int_0^{t'} dt_2 \langle F(t_1)F(t_2)\rangle_{l,L}\right],$$

where

$$\langle F(t_1)F(t_2)\rangle_{l,L}$$
$$= \xi^2\nu^2\{[\bar{n}e^{i\nu(t_1-t_2)} + (\bar{n}+1)e^{-i\nu(t_1-t_2)}]e^{-\alpha|t_1-t_2|} \quad (18)$$
$$+ 2(\bar{n}_o - \bar{n})\cos\nu(t_1-t_2)e^{-\alpha(t_1+t_2)}\}$$

$$+ \sum_k \lambda_k^2 [\bar{n}_k e^{i\omega_k(t_1 - t_2)} + (\bar{n}_k + 1)e^{-i\omega_k(t_1 - t_2)}],$$

where $\bar{n}_k = [\exp(\omega_k/k_B T) - 1]^{-1}$, $\bar{n}_o = [\exp(\nu/k_B T_o) - 1]^{-1}$ is the occupation number for the local mode in a nonequilibrium state at the "spin" temperature $T_o$, which differs from the lattice temperature $T$; in what follows, however, we neglect this temperature difference assuming that $\bar{n}_o = \bar{n}$.

The steady-state form of the correlation function (17), with allowance for Eq. (18) and for the usually satisfied inequality $\alpha/\nu \ll 1$, acquires the form

$$ln\langle U^+(t) U(t')\rangle_{l,L}$$
$$\approx \xi^2\{-(2\bar{n} + 1) - [(2\bar{n} + 1)\alpha + i\nu](t - t')$$
$$+ [\bar{n} e^{i\nu(t - t')} + (\bar{n} + 1)e^{-i\nu(t - t')}]e^{-\alpha|t - t'|}\} \quad (19)$$
$$+ \sum_k \left(\frac{\lambda_k}{\omega_k}\right)^2 [-(2\bar{n}_k + 1) - i\omega_k(t - t')$$
$$+ \bar{n}_k e^{i\omega_k(t - t')} + (\bar{n}_k + 1)e^{-i\omega_k(t - t')}].$$

We will further omit the coefficients $i\xi^2\nu$ and $i\sum_k \lambda_k^2/\omega_k$ in the terms linear in $(t - t')$ on the right-hand side of Eq. (19), assuming them to be already included in the renormalized energies of the ground and excited electronic states.

For $t \gg \alpha^{-1}, k^{-1}$, we may use the Markovian approximation by passing, in Eq. (15), to integration over $(t - t')$ and replacing the upper integration limit by infinity. Then, Eq. (15) gains the form [19, 21]

$$\frac{\partial\sigma(t)}{\partial t} = -\pi g^2 F(\omega; T)[R^+R^-\sigma(t)$$
$$- 2R^-\sigma(t)R^+ + \sigma(t)R^+R^-] + \Lambda_a\sigma(t), \quad (20)$$

where we introduced the normalized function of spectral distribution

$$F(\omega; T) = \frac{1}{2\pi}\int_{-\infty}^{\infty} d\tau e^{i(\varepsilon - \omega)\tau - ks|\tau|}\langle U^+(\tau)U(0)\rangle_{l,L},$$
$$\int_{-\infty}^{\infty} d\omega F(\omega; T) = 1. \quad (21)$$

On account of Eq. (19), function (21) may be represented in the form

$$F(\omega; T) = \frac{1}{2\pi} e^{-\xi^2(2\bar{n} + 1)} \sum_{p = -\infty}^{\infty} \sum_{r, s = 0}^{\infty} \prod_k \frac{\xi^{2(r+s)}(\bar{n} + 1)^r \bar{n}^s 2[\kappa + (r + s)\alpha + \xi^2(2\bar{n} + 1)\alpha]}{r!s![\varepsilon - \omega - \nu(r - s) - p\omega_k]^2 + [\kappa + (r + s)\alpha + \xi^2(2\bar{n} + 1)\alpha]^2}$$
$$\times e^{-w(T)} I_p(\lambda_k/\omega_k; T)\left(\frac{\bar{n}_k + 1}{\bar{n}_k}\right)^{p/2}, \quad (22)$$

where $e^{-w(T)}$ is the Debye–Waller factor,

$$w(T) \equiv \sum_k \left(\frac{\lambda_k}{\omega_k}\right)^2 (2\bar{n}_k + 1), \quad (23)$$

$I_p$ is the Bessel function of the imaginary argument of the first kind. Note that

$$I_p(\lambda_k/\omega_k; T) \equiv I_p\left(2\frac{\lambda_k^2}{\omega_k^2}\sqrt{\bar{n}_k(\bar{n}_k + 1)}\right). \quad (24)$$

It is worthwhile to make a few remarks about the form of function (22). First, we will assume that the interaction of the light with the optical center occurs in such a way that the number of the lattice phonons before and after the interaction is the same (elastic process), i.e., only one term with $p = 0$ should be retained in the sum over $p$. Second, to be able to study the processes of spontaneous emission with the participation of a specified number of created and annihilated quanta of the local vibration, we may use, in Eq. (22), the resonance approximation by removing summation over $r$ and $s$. Then, the function

$$F^{(s, r)}(\omega = \varepsilon - \nu(r - s); p = 0; T) \equiv F_o^{(s, r)}(T)$$
$$= \frac{1}{\pi} e^{-\xi^2(2\bar{n} + 1)} \frac{\xi^{2(r+s)}(\bar{n} + 1)^r \bar{n}^s}{r!s![\kappa + (r + s)\alpha + \xi^2(2\bar{n} + 1)\alpha]} \quad (25)$$
$$\times e^{-w(T)}\prod_k I_0(\lambda_k/\omega_k; T)$$

determines the contribution to emission of the light quantum with the frequency $\omega = \varepsilon - \nu(r - s)$ on the vibronic transition from the $s$th vibrational level in the excited electronic state ($s$ vibrational quanta with the frequency $\nu$ are annihilated) to the $r$th vibrational level of the ground electronic state ($r$ vibrational quanta of the same frequency $\nu$ are created) with an unchanged vibrational state of the host lattice. Therefore, the master equation for the density matrix, describing the $es \longrightarrow gr$ vibronic transition (where $e$ and $g$ are the excited and ground electronic states) for unchanged occupation numbers of the lattice modes, as follows from Eq. (20), should be represented in the form

$$\frac{\partial \sigma^{(s,r)}(t)}{\partial t} = -\pi g^2 F_o^{(s,r)}(T)[R^+R^-\sigma^{(s,r)}(t) \\ - 2R^-\sigma^{(s,r)}(t)R^+ + \sigma^{(s,r)}(t)R^+R^-] + \Lambda_a \sigma^{(s,r)}(t). \quad (26)$$

## PARAMETERS OF VIBRONIC SUPERRADIANCE

The emission intensity $I^{(s,r)}$ of the $es \longrightarrow gr$ vibronic transition is determined by the relationship

$$I^{(s,r)}(t) = -[\varepsilon - \nu(r-s)]\frac{\partial}{\partial t}\text{Sp}_a(R^z\sigma^{(s,r)}(t)), \quad (27)$$

which, for the initially totally inverted system of $N$ optical centers with no dephasing (neglecting the term $\Lambda_a \sigma$), yields

$$I^{(s,r)}(t) = \frac{N[\varepsilon - \nu(r-s)]}{4\tau_R^{(s,r)}(T)}\text{sech}^2\left(\frac{t - t_D^{(s,r)}(T)}{2\tau_R^{(s,r)}(T)}\right). \quad (28)$$

Characteristic parameters of the vibronic superradiance in Eq. (28), namely, the duration of the emission $\tau_R^{(s,r)}$ and the delay time of the superradiance pulse $t_D^{(s,r)}$, are determined by the formula

$$[\tau_R^{(s,r)}(T)]^{-1} = 2\pi N g^2 F_o^{(s,r)}(T), \\ t_D^{(s,r)}(T) = \tau_R^{(s,r)}(T)\ln N, \quad (29)$$

where the function $F_o^{(s,r)}(T)$ is defined by Eq. (25).

As seen from Eqs. (28), (29), and (25), the higher the spectral density $F_o^{(s,r)}$ of the $es \longrightarrow gr$ vibronic transition, the shorter the emission duration and delay time and the higher the intensity of the superradiance pulse generated on this transition. For all vibronic transitions, the spectral density function $F_o^{(s,r)}$ decreases with increasing Stokes losses $\xi^2(2\bar{n} + 1)$ per a local mode and Stokes losses $\sum_k \frac{\lambda_k^2}{\omega_k^2}(2\bar{n}_k + 1)$ summed over all the lattice modes through the relevant Debye–Waller factors $\exp[-\xi^2(2\bar{n} + 1)]$ and $\exp[-w(T)]$. For $k_B T \ll \nu$, the dependence of the spectral density distribution and, hence, of the superradiance parameters on temperature is predominantly controlled by the Debye–Waller factor of the lattice modes and by the product of the zero-order Bessel functions $\prod_k I_0(\lambda_k/\omega_k; T)$, since $\omega_k \ll \nu$ and, under the conditions of the experiment, $\bar{n}_k(T) \sim O(1)$ while $\bar{n}(T) \sim o(1)$.

For comparison with the experimental data, it may be useful to know the ratio of the delay times and peak intensities for different vibronic transitions. From Eqs. (28), (29), and (25), assuming that $\bar{n} \approx 0$, we have

$$\frac{t_D^{(s,r)}}{t_D^{(s',r')}} = \frac{r!s![k + (r+s)\alpha + \xi^2\alpha]}{r'!s'![k + (r'+s')\alpha + \xi^2\alpha]}\xi^{2(s'+r'-s-r)}, \quad (30)$$

$$\frac{I_{\max}^{(s,r)}}{I_{\max}^{(s',r')}} = \frac{\varepsilon - \nu(r-s)}{\varepsilon - \nu(r'-s')}\frac{t_D^{(s,r)}}{t_D^{(s',r')}}. \quad (31)$$

These relationships do not depend on temperature because the temperature factor, related to the lattice modes, is canceled in the ratios, and the local mode cannot be thermally excited at $k_B T \ll \nu$.

## TWO-COLOR SUPERRADIANCE. COMPARISON WITH THE EXPERIMENT

In the experiments [7, 8], the $^2\Pi_g$–$^2\Pi_u$ electronic transition of the molecular ions $O_2^-$ in KCl was excited by 30-ps pulses at the wavelength 266 nm. For the pump power density exceeding 20 GW/cm$^2$ and at low temperatures ($T < 30$ K), a two-color (yellow–red) coherent emission on the vibronic transitions 0–10 and 0–11 (with the wavelengths 592.8 and 629.1 nm) was observed. The intensity of the emission exceeded by a factor of $10^4$ the usual spontaneous emission. When the pump power decreased by about a factor of two, down to 10 GW/cm$^2$, the yellow superradiance pulse disappeared and only the red one could be observed. The measured delay time of the pulses varied from 0.5 to 10 ns. The spot of the focused pump was 0.01 cm in diameter, and the initial inversion density, for the sample length $l = 1$ cm, was estimated to be $n = N/V = 10^{17}$ cm$^{-3}$ (where the volume $V$ was $0.01 \times 0.01 \times 1 = 10^{-4}$ cm$^3$), though this value, as was pointed out in [39], may be slightly overestimated. The homogeneous and inhomogeneous relaxation times were $T_2 = 100$ ps (at $T = 4.2$ K) and $T_2^* = 30$ ps, respectively. The radiative decay time was $T_1 = 10$ ns.

First of all, we may compare the theoretical and experimental ratios of the delay times of the yellow and red superradiance pulses. From Eq. (30) we have

$$\frac{t_D^{(0,10)}}{t_D^{(0,11)}} = \frac{\xi^2[k + (10 + \xi^2)\alpha]}{11[k + (11 + \xi^2)\alpha]} \approx 0.78 \quad (32)$$

(here we take into account that, for $O_2^-$ in KCl, the dimensionless Stokes losses are $\xi^2 = 9$, i.e., the 0–9 vibronic line shows the highest oscillator strength [8], the decay time of the vibrational levels is $\alpha^{-1} = 50$ ps [7], and $c/l = 3 \times 10^{10}$ s$^{-1}$). Thus, for the pump power exceeding the threshold (20 GW/cm$^2$), the yellow pulse should be generated first and the red one, second. Indeed, it was found experimentally (see Fig. 7 in [8]) that the statistically averaged delay time of the yellow pulse is smaller than that of the red one approximately in the proportion predicted by theoretical estimates. The ratio of the peak intensities calculated using Eqs. (30)–(32)

$$\frac{I_{\max}^{(0,10)}}{I_{\max}^{(0,11)}} = \frac{\varepsilon - 10\nu}{\varepsilon - 11\nu} \frac{t_D^{(0,11)}}{t_D^{(0,10)}} = 1.36, \qquad (33)$$

also corresponds to the experimental data (see Fig. 7 in [8]). It is also clear why the red pulse is observed experimentally at lower pump power than the yellow pulse: on the one hand, $t_D^{(0,10)}$ and $t_D^{(0,11)} \sim 1/n$, and an increase of $n$ evidently corresponds to an increase in the pump power, and; on the other hand, as is dictated by relationship (32), $t_D^{(0,11)} > t_D^{(0,10)}$ [as well as, according to (29), $\tau_D^{(0,11)} > \tau_D^{(0,10)}$, and, therefore, as the pump power increases, the required values are first reached for the larger parameters of the superradiance [$t_D^{(0,11)}$ and $\tau_D^{(0,11)}$], i.e., the red pulse is initiated first.

The absolute value of the time scale unit of the superradiance may be found using the Friedberg–Hartmann relation [39]

$$\kappa l = \frac{T_2'}{\tau_R}, \qquad (34)$$

where $\kappa$ is the linear gain factor and $T_2' = (1/T_2 + 1/T_2^*)^{-1}$. Since in experiments [7, 8], the superradiance signal was amplified by a factor of $10^3$, then, using the Beer law, we have the estimate $\kappa l \approx 9.2$, whence, by formula (34), we find the time scale $\tau_R \approx 2.5$ ps. On the other hand, by comparing the characteristic value of $\tau_R$ with the superradiance times, determined by relationships (29) for each vibronic transition, we may find the density of the particles needed for implementation of the superradiance effect. Using the above relaxation parameters and the force characteristics of the interaction of the $O_2^-$ centers with the radiation field and intramolecular vibrations, we, at first, have

$$\tau_R^{(0,11)} = \left[4.5 \times 10^{-5} n e^{-w(T)} \prod_k I_0(\lambda_k/\omega_k; T)\right]^{-1}, \qquad (35)$$

$$\tau_R^{(0,10)} = 0.78 \tau_R^{(0,11)}, \qquad (36)$$

hence, by equating $\tau_R^{(0,11)}$ to the time scale unit of the superradiance $\tau_R = 2.5$ ps, we obtain the functional relation between the density of the superradiating particles, proportional to the pump power, and the temperature of the crystal, its dynamic characteristics, and Stokes losses into the lattice modes:

$$n = \frac{N}{V} = 0.9 \times 10^{16} e^{w(T)} \left[\prod_k I_0(\lambda_k/\omega_k; T)\right]^{-1}. \qquad (37)$$

To estimate the temperature factor in Eqs. (35) and (37), we approximate the phonon spectrum of KCl in the low-temperature range by the effective frequency $\bar{\omega} \approx 40$ cm$^{-1}$ and take the Stokes losses into the lattice modes to be equal to 0.23 [31]. Then, for $T = 4.2$ K, using Eq. (37), we obtain the initial concentration $n = 1.2 \times 10^{16}$ cm$^{-3}$, which should enter into formula (35) for $\tau_R^{(0,11)}$. To implement the superradiance on the 0–10 vibronic transition, the initial density $n$ of the superradiating particles (or, which is the same, the pump power) should be increased by approximately 30% to meet condition (36). In the experiments [7, 8], the transition from single-color to two-color superradiance required nearly a twofold enhancement of the pump power. A possible reason for this is that the theoretical estimate does not take into account a deeper correlation in the formation of the superradiance pulses on different vibronic transitions, since, when going from Eq. (20) to Eq. (26), we used the "resonance" approximation. On the other hand, the condition $t_D^{(0,10)}, t_D^{(0,11)} < T_2'$, needed to observe the "pure" superradiance [40], was not satisfied in the experiment.

As the temperature increases, the Debye–Waller factor decreases faster than the zero-order Bessel function increases. According to Eqs. (35), (36), and (29), this leads to an increase of duration and delay times for the superradiance pulses. The increase of the superradiance time, in turn, reduces the gain $\kappa l$ (34). Note that, in conformity with the Beer law, a decrease in the gain by a factor of $m$ corresponds to a decrease in the intensity by $m$ orders of magnitude. The superradiance intensity on the vibronic transitions 0–10 and 0–11 measured experimentally [7, 8] exceeded the spontaneous background by a factor of $10^4$. Thus it is clear that an increase in the superradiance pulse duration, e.g., by a factor of two, due to the increasing temperature of the sample, should suppress nearly completely the superradiance effect. Indeed, the superradiance could be observed only below some critical temperature which depended on concentration of the $O_2^-$ centers, namely, for the highest achievable concentration $7 \times 10^{17}$ cm$^{-3}$, this temperature was 27 K and was reduced to 8 K for the concentration $8 \times 10^{16}$ cm$^{-3}$. Thus, to compensate for the destructive effect of the thermal factor, one has, when increasing the sample temperature, to increase the concentration of the optical centers (or the pump power), which also follows from Eq. (35). Formula (37), which connects concentration of the optical centers and crystal temperature in such a way that the

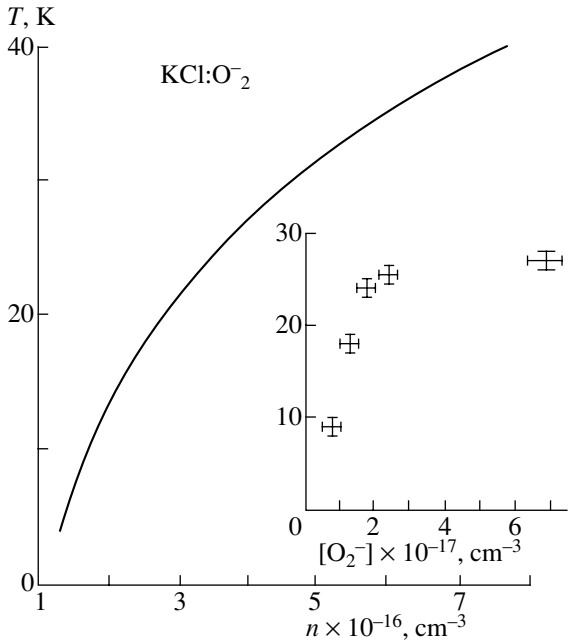

The relationship between the concentration of the optical impurities $n$ and the crystal temperature $T$, for which the superradiance may be still observed. The experimental data [8] are shown in the inset.

superradiance time remains fixed, may be used to account for (at least qualitatively) the critical temperature dependence of the superradiance effect. The curve in the figure shows (the experimental dependence [8] is shown in the inset of the figure) how the concentration of the optical centers should be increased with the increasing temperature of the sample so as not to destroy conditions for the implementation of the superradiance effect.

It is noteworthy that two-color superradiance has recently been observed experimentally [41] in the Van Vleck paramagnetic $LaF_3$: $Pr^{3+}$, but on the pure electronic, rather than on vibronic, transitions $^3P_0$–$^3H_4(0)$ and $^3P_0$–$^3H_6$.

## CONCLUSION

The master equation for the Dicke superradiance on vibronic transitions of impurity crystals is derived (see also [42, 43]). The superradiance parameters (the delay times, durations and intensities of the superradiance pulses) for each vibronic transition are shown to depend on the strength of coupling between the electronic states and the intramolecular impurity mode that produces the vibronic structure of the optical spectrum in the form of vibrational sidebands of the pure electronic line, and on the crystal temperature via the Debye–Waller factor of the lattice vibrations. The theoretical estimates of the ratio of the delay times and intensities of the superradiance pulses for different vibronic transitions well agree with the experimental data on observation of the two-color superradiance in the polar dielectric $KCl:O_2^-$. In addition, the theory describes qualitatively well the critical temperature dependence of the superradiance.